\documentclass[12pt]{JHEP3}
\usepackage{amsmath}

\newcommand{\be}{ \begin{equation}}
\newcommand{\ee}{\end{equation}}
\newcommand{\bea}[1]{\begin{eqnarray}\label{#1} }
\newcommand{\eea}{\end{eqnarray}}

\def\ZZZ{{\hskip-3pt\hbox{ Z\kern-1.6mm Z}}}
\def\zzz{{\hskip-3pt\hbox{ z\kern-1mm z}}}

\newcommand{\W}{{\cal W}}
\newcommand{\f}{{\rm f}}

\newcommand{\half}{{1\over 2}}

\def\one{{\hbox{ 1\kern-.8mm l}}}
\def\zero{{\hbox{ 0\kern-1.5mm 0}}}

\newcommand{\hs}[1]{\mbox{hs$[#1]$}}
\newcommand{\w}[1]{\mbox{$\mathcal{W}_\infty[#1]$}}

\title{Triality in Minimal Model Holography}

\author{
Matthias R.\ Gaberdiel$^{a}$ and Rajesh Gopakumar$^{b}$ \\ 
$^a$Institut f\"ur Theoretische Physik, ETH Zurich, \\
$\;$CH-8093 Z\"urich, Switzerland \\
$\;$\email{gaberdiel@itp.phys.ethz.ch}\\ \\ 
$^b$Harish-Chandra Research Institute, \\
$\;$Chhatnag Road, Jhusi,\\
$\;$Allahabad, India 211019\\
$\;$\email{gopakumr@hri.res.in}}

\abstract{The non-linear $\W_{\infty}[\mu]$ symmetry algebra underlies the 
duality between the $\W_N$ minimal model CFTs and the $\hs{\mu}$ higher spin theory 
on AdS$_3$. It is shown how the structure of this symmetry algebra at the {\it quantum} level,
i.e.\ for finite central charge, can be determined completely. The resulting algebra exhibits
an exact equivalence (a `triality') between three (generically) distinct values of the 
parameter $\mu$. This explains, 
among other things, the agreement of symmetries between the $\W_N$ minimal 
models and the bulk higher spin theory. We also study the consequences of this triality for 
some of the simplest $\W_{\infty}[\mu]$ representations, thereby clarifying 
the analytic continuation between the `light states' of the minimal models and conical 
defect solutions in the bulk. These considerations also lead us to propose that 
one of the two scalar  fields in the bulk actually has a non-perturbative origin.}

\preprint{HRI-P-12-05-001}

\begin{document}

\section{Introduction and Summary}

Symmetry plays a very powerful role in the AdS/CFT correspondence. The presence of 
large symmetries in both the bulk and the boundary can, in some instances, effectively constrain 
the dynamics so that the equivalence between the two descriptions is largely a consequence of 
the matching of the symmetries. Such examples, in turn, can help in deciphering the holographic
dictionary better. The developments which have uncovered the planar integrability of 
${\cal N}=4$ Super Yang-Mills theory and the related integrability of the string sigma model 
on AdS$_5\times $S$^5$ go in this direction. Interestingly, these enlarged symmetries are 
usually not very manifest and their matching on both sides is a nontrivial fact. 

The presence of supersymmetry is usually a necessary prerequisite for these larger 
symmetries, and most of the well studied examples of AdS/CFT exploit the power of 
supersymmetry. 
It has gradually been realised that higher spin symmetries can play an analogous role in 
effectively constraining the dynamics in non-supersymmetric contexts. These symmetries 
might essentially govern the vicinity of the tensionless limit of string theory on AdS, or 
equivalently the weak coupling limit of gauge theories. 
Even when the symmetries are (mildly) broken they can provide constraints on the form of 
the correlation functions \cite{Maldacena:2011jn,Maldacena:2012sf}. While for
CFTs in $d>2$ higher spin symmetries can only be realised exactly in free boson/fermion
theories,\footnote{This conclusion relies on a number of general assumptions, one of
which is that the number of degrees of freedom, $N$, is finite.} in $d=2$
this conclusion may be evaded (as in the Coleman-Mandula theorem). 
Indeed, there is a large class of 
interacting two dimensional CFTs which have (holomorphic) conserved currents of arbitrarily 
high spin. These symmetries usually define so-called $\W$-algebras. They
are generically nonlinear in the sense that the generators do not form a conventional
Lie algebra, but that the commutators of two generators can only be expressed in terms
of quadratic (or even higher order) products of the generators.

An important step was taken in \cite{Henneaux:2010xg, Campoleoni:2010zq}, where it was 
realised that the asymptotic symmetry algebras of higher spin theories on AdS$_3$ are such 
$\W$-algebras. The particular cases studied in \cite{Henneaux:2010xg, Campoleoni:2010zq}
actually belong to a one-parameter family of higher spin  theories whose symmetry algebras
could be identified with  $\W_{\infty}[\mu]$  \cite{Gaberdiel:2011wb} 
(see also \cite{Campoleoni:2011hg}). 
These generalisations of the Brown-Henneaux result 
opened the possibility of a 2d CFT with $\W$-symmetry being dual to some higher spin theory 
on AdS$_3$. A concrete proposal was then made in \cite{Gaberdiel:2010pz},
relating the so-called $\W_{N,k}$ family of unitary CFTs (in a certain large $N$, $k$ 't~Hooft 
limit) to a specific higher spin theory (coupled to additional scalar fields) on AdS$_3$. This 
duality has been further investigated in
\cite{Gaberdiel:2011zw,Chang:2011mz,Papadodimas:2011pf,Ahn:2011by,Castro:2011iw,%
Ammon:2011ua,Chang:2011vk,Vasiliev:2012vf}.

However, the symmetries on the bulk and boundary theories are not obviously the same, 
though they are both $\W$-algebras. At any fixed $N$, 
the minimal model CFTs have $\W_N$ symmetry, which corresponds 
to $\W_{\infty}[\mu]$ with $\mu=N$; at these values the algebra truncates consistently to 
one with currents of a maximum spin $s=N$. On the other hand, the bulk theory is based 
on the $\hs{\lambda}$ higher spin theory with $\lambda$ identified with the 't~Hooft coupling 
$\lambda=\tfrac{N}{N+k}$, and its asymptotic symmetry algebra is
$\W_{\infty}[\mu]$ with $\mu=\lambda$ \cite{Gaberdiel:2011wb}.
At first sight this appears to be rather different from the symmetries of a theory with 
$\mu=N$. 

The crucial point, however, is that the Brown-Henneaux like analysis of 
\cite{Henneaux:2010xg, Campoleoni:2010zq, Gaberdiel:2011wb} is `classical', i.e.\ 
it determines the Poisson brackets of the $\W$-generators, and is 
only valid at large $c=\tfrac{3\ell}{2G_N}$; in order to emphasize this aspect, we shall
sometimes denote this {\it classical} $\W_{\infty}[\mu]$ algebra by $\W^{\rm cl}_{\infty}[\mu]$.
When the central charge is finite and Poisson brackets are replaced by commutators, 
the non-linear nature of $\W_{\infty}[\mu]$ leads to additional terms arising from the normal
ordering of products and the requirement to satisfy the Jacobi identities. As a consequence, 
the {\it quantum}  $\W_{\infty}[\mu]$ algebra, $\W^{\rm qu}_{\infty}[\mu]$, is a significant 
deformation of the classical algebra. 

Therefore, while the classical algebras $\W^{\rm cl}_{\infty}[\mu]$ for $\mu=N$ and 
$\mu=\lambda$ are certainly not isomorphic, this does not preclude that there exists a non-trivial
equivalence of the corresponding quantum $\W$-algebras. In fact, heuristic 
considerations \cite{Gaberdiel:2011zw} based on generalised level-rank dualities for coset 
CFTs \cite{Kuniba:1990zh, Altschuler:1990th} suggest such a relation; this 
will be reviewed at the beginning of section~2 below. 
In this paper, we put this equivalence on a firmer footing by giving compelling evidence for a 
general triad of isomorphic $\W^{\rm qu}_{\infty}[\mu]$ algebras for (three generically) different 
values of $\mu$. As we will see, this triality implies the desired equivalence in the case 
when  one of the values of $\mu$ is $\mu=N$. 

More specifically, as we shall explain in section~2.1, we can determine the structure of the 
quantum $\W^{\rm qu}_{\infty}[\mu]$ algebra completely, using two constraints. 
Starting from the classical
algebra  $\W^{\rm cl}_{\infty}[\mu]$, the requirement that the Jacobi identities are satisfied
fixes the correct form of the normal ordered products, as well as the finite shifts
in the coefficients of the non-linear terms. This determines the algebra up to the form
of some structure constants that are only known from $\W^{\rm cl}_{\infty}[\mu]$ to
leading order in $\tfrac{1}{c}$. The complete $c$-dependence of these structure constants can
then be determined by requiring that the representation theory of $\W^{\rm qu}_{\infty}[\mu]$
matches that of $\W_N$ for $\mu=N$. In general, the resulting quantum algebra does 
does not actually contain $\hs{\mu}$ as a subalgebra;  instead $\hs{\mu}$ is only 
a subalgebra in the $c\rightarrow \infty$ limit.

As it turns out, the quantum algebra $\W^{\rm qu}_{\infty}[\mu]$ is
more invariantly parametrised in terms of two numbers: these are $c$, the central charge, 
and $\gamma$, the structure constant 
which captures the leading nontrivial higher spin coupling (of the spin four current in the 
OPE of two spin three currents). All other structure constants appear to be 
fixed in terms of these
two parameters. Furthermore, $\mu$ is determined by a cubic equation which depends 
only on $(c,\gamma)$, and therefore there are three values of $\mu$ which correspond 
to isomorphic algebras. This effectively proves the quantum equivalence of these
three $\W^{\rm qu}_{\infty}[\mu]$ algebras.
\smallskip

In section~3 we then go on to study some of the simplest representations of 
$\W^{\rm qu}_{\infty}[\mu]$ which we call minimal representations. These are the 
representations which have the fewest number of low-lying states. It turns out that there 
are three of them for fixed values of $(c,\gamma)$. We verify 
that the quantum numbers (dimensions, low spin charges) of these three representations are
indeed consistent with the above triality.\footnote{Interestingly, one of the three is not a 
representation of $\hs{\mu}$, even in the large $c$ limit.} For the case of the $\W_{N,k}$ 
minimal models, the two physical 
representations correspond to the basic coset primaries labelled as $(0;\f)$ and $(\f;0)$ 
(and their complex conjugates). The fusion of these two representations produces other 
non-minimal representations such as $(\f;\f)$ --- the lightest of the light states (in the large 
$N$ 't~Hooft limit).  

Therefore, in section~4 we revisit these light states and their relation to semi-classical 
solutions (conical defects) \cite{Castro:2011iw} of the bulk ${\rm SL}(N)$ higher spin theory. 
Since our analysis of the minimal representations and their fusion holds for all values of 
$c$, we can study the representation theory at 
fixed $N$ (which then determines 
$\gamma =\gamma(N,c)$), and hence understand the behaviour of the various
representations as a function of $c$. 
In particular, we consider the analytic continuation of the $(\f;\f)$ state from 
$c=c_{N,k} \leq (N-1)$ (the value for the minimal models),  to the semi-classical regime
$c\rightarrow \infty$. In the latter regime, its quantum 
numbers match (to leading order in $c$) those of the conical defect solution of the bulk theory.
We can similarly look at all the other light states $(\Lambda; \Lambda)$ of the CFT and 
continue the representations (at fixed $N$) to large $c$. The expressions for charges and 
dimensions are smooth functions of $c$, and their leading behavior matches with those 
calculated for the conical defects \cite{Castro:2011iw}. This shores up the identification of the 
latter with the light states in a precise analytic continuation in $c$ of representations of 
$\W^{\rm qu}_{\infty}[\mu]$. 

If we consider the similar analytic continuation in $c$ (again at fixed $N$) 
of the two minimal representations $(0;\f)$ and $(\f;0)$, we find that the dimension of the former is 
proportional to $c$ (for large $c$) while that of the latter is of order one. Thus the two
states are on a different footing, and it  appears to be more natural to consider the former 
as non-perturbative (or solitonic), whereas the latter can be viewed as a perturbative 
excitation. 
Therefore, in section~5  we are led to refine the conjecture of  \cite{Gaberdiel:2010pz},
and propose that the bulk theory should be considered to be the $\hs{\lambda}$ theory 
with only {\it one} complex scalar (with $m^2=-1+\lambda^2$), and quantized in the standard
way (the $+$ quantization). The other primary corresponding to $(0;\f)$ is to be viewed 
as an excited state of the lightest conical defect. We believe this alternative picture 
explains some of the puzzling aspects of the light states and their relation to the 
perturbative excitations. We summarise the current status of the duality and interesting
avenues for further work in section~6.

\section{The Structure of the $\w{\mu}$-Algebra}\label{sec:walgebra}

Let us begin by motivating why there should be non-trivial identifications among 
the quantum $\w{\mu}$ algebras. Recall that by construction, $\W_N$ agrees with
$\w{\mu}$ for integer $\mu=N$. Indeed 
$\w{\mu}$  is the Drinfeld-Sokolov reduction of
$\hs{\mu}$,  and $\hs{\mu}$ reduces\footnote{Strictly speaking,
the relation is that $\hs{N}$ contains a large ideal, and that the quotient of 
$\hs{N}$ by this ideal is equivalent to $\mathfrak{su}(N)$. We will come back to this
in section~\ref{finiteNsub}.} to $\mathfrak{su}(N)$  for $\mu=N$, whose
Drinfeld-Sokolov reduction is $\W_N$; thus we have the relation
\be\label{obvious}
\left. \w{\mu} \right|_{\mu=N} \cong {\cal W}_N  \ .
\ee
However, there is also a second, somewhat more subtle, relation between $\w{\mu}$ 
and ${\cal W}_N$.  It was conjectured in 
\cite{Kuniba:1990zh,Altschuler:1990th} that the coset models
\be\label{cosetdef}
{\cal W}_{N,k} \equiv
\frac{\mathfrak{su}(N)_k \oplus \mathfrak{su}(N)_1}{\mathfrak{su}(N)_{k+1}} \ \cong \ 
\frac{\mathfrak{su}(M)_l \oplus \mathfrak{su}(M)_1}{\mathfrak{su}(M)_{l+1}}   \equiv 
{\cal W}_{M,l}
\ee
are related to one another by some sort of level-rank duality, where 
the relation between the parameters is 
\be\label{rel2}
k = \frac{N}{M} - N \ , \qquad l = \frac{M}{N} - M \ .
\ee
Here $M$ and $N$ are taken to be positive integers, whereas $k$ and $l$ are fractional 
(real) numbers, and the central charges of both sides are equal to 
\be\label{cNk}
c_{N,k} \equiv (N-1) \, \Bigl[ 1 - \frac{N (N+1)}{(N+k) (N+k+1)} \Bigr]  = 
(M-1) \Bigl[ 1 - \frac{M (M+1)}{(M+l) (M+l+1)} \Bigr]  \equiv  \, c_{M,l} \ .
\ee
However, it seems reasonable to assume that this level-rank duality will 
also hold if instead of integer $N$, $M$, we consider the situation where $N$ and $k$
are integers. Then we can solve (\ref{rel2}) for $M$ to obtain
\be
M \equiv \lambda = \frac{N}{N+k} \ ,
\ee
while $l$ is determined by the condition that both sides have the same central charge. 
Next we observe that we have also quite generically that
\be
\frac{\mathfrak{su}(M)_l \oplus \mathfrak{su}(M)_1}{\mathfrak{su}(M)_{l+1}}  \ \cong \ 
\hbox{Drinfeld-Sokolov reduction of} \ \mathfrak{su}(M)\ \hbox{at level ${\hat{l}}$} \ , 
\ee
where again $\hat{l}$ is determined so as to have the same central charge as the left-hand-side.
For non-integer $M$ we can think of 
\be
\mathfrak{su}(\lambda) \cong \hs{\lambda} \ ,
\ee
and the Drinfeld-Sokolov reduction of $\hs{\lambda}$ equals $\w{\lambda}$. Combining these statements
then leads to the claim that we have an isomorphism of algebras
\be\label{claim}
{\cal W}_{N,k} \equiv
\frac{\mathfrak{su}(N)_k \oplus \mathfrak{su}(N)_1}{\mathfrak{su}(N)_{k+1}} \ \cong \ 
\w{\lambda}   \qquad \hbox{with} \ \lambda = \frac{N}{N+k} \ .
\ee
Here the central charge of $\w{\lambda}$ is taken to agree with that of ${\cal W}_{N,k}$,
i.e.\ with $c_{N,k}$ defined in (\ref{cNk}). 
This relation should be true not just in the 't~Hooft limit, but also for finite $N$ and $k$
(and hence finite central charge).

Actually, there is a second variant of this relation. The ${\cal W}_N$ algebra at level $k$
is identical to the ${\cal W}_N$ algebra at level 
\be\label{khat}
\hat{k} = - 2 N - k - 1 
\ee
since the central charges of the two algebras agree, i.e. $c_{N,k} = c_{N,\hat{k}}$. 
Incidentally, this identification has a natural interpretation from the Drinfeld-Sokolov (DS)
point of view. Recall that the cosets $\W_{N,k}$ in (\ref{cosetdef}) 
are equivalent to the DS reduction of $\mathfrak{su}(N)$ at level $k_{\rm DS}$, where 
the two levels are related as (see e.g.\ \cite{BouScho93} for a review of these matters)
\be\label{DSrel}
\frac{1}{k+N} = \frac{1}{k_{\rm DS}+N} - 1 \ .
\ee
{}From the DS point of view, replacing $k\mapsto \hat{k}$ as in (\ref{khat}) is equivalent
to replacing  $k_{\rm DS}$ by $\hat{k}_{\rm DS}$ with 
\be
\hat{k}_{\rm DS} + N  = \frac{1}{k_{\rm DS}+N}  \ .
\ee
In terms of the underlying free field description, this corresponds to exchanging
(see e.g.\ \cite{BouScho93} or \cite[Section 6.2.2]{Gaberdiel:2011zw}) the roles of 
$\alpha_\pm$, i.e.\ to define 
$(\hat\alpha_+,\hat\alpha_-) = (-\alpha_-,-\alpha_+)$. 
This is an obvious symmetry of the DS reduction under which the representations
are related as $\Lambda_+\leftrightarrow \Lambda_-^\ast$, see also the discussion
in section~\ref{sec:anaper}.

Thus we can repeat the above analysis with $\hat{k}$ in place of $k$, to conclude that 
$\W_{N,k}$ is also equivalent to $\w{\mu}$ with $\mu=- \tfrac{N}{N+k+1}$. 
Altogether this suggests that we have the `triality' 
\be\label{tria1}
\w{N} \ \cong \ \w{\frac{N}{N+k}} \  \cong \ 
\w{-\frac{N}{N+k+1}} \qquad
\hbox{at $c=c_{N,k}$.}
\ee
In the following we want to give highly non-trivial evidence for this claim. (Actually,
as we shall see, a somewhat stronger statement appears to be true in that
we need not even assume that $N$ is integer.)  In order to discuss these issues, however,
we first need to understand the explicit structure of $\w{\mu}$ in more detail.

\subsection{Explicit Form of the Algebra}

We can derive the commutation relations of the quantum $\w{\mu}$ algebra 
by starting with the classical  $\w{\mu}$ algebra
that can be defined as the asymptotic symmetry algebra
of Chern-Simons theory based on $\hs{\mu}$.
The finite $c$ corrections to the non-linear terms can then be determined recursively 
by solving the Jacobi identities. Using the results of \cite{Gaberdiel:2011wb} (see also 
\cite{GHJ}) we have worked this out explicitly for the first few terms, and the
resulting commutation relations are given in appendix~\ref{app:comm}. To the
order to which we have studied this problem, these considerations fix the entire structure of 
the commutators completely, except for the 
full $c$-dependence of the structure constant $C^4_{33}$.
Schematically, this is the structure constant appearing in the OPE
\be\label{c334}
W\cdot W \sim C^4_{33}U +\cdots \ ,
\ee
where $W,U$ are the spin three and spin four currents respectively. 
It follows from the analysis of appendix~\ref{app:comm} (see eq.~(\ref{ww})) that 
\be\label{c334c}
C^4_{33} = 8\, \sqrt{\tfrac{1}{5}\, \tfrac{\mu^2-9}{\mu^2-4}}  
+ {\cal O}\left( \tfrac{1}{c} \right) \ . 
\ee
In order to determine the full $c$ dependence of this structure constant, we can study
the representation theory of the resulting algebra, and demand that it is compatible, 
for $\mu=N$, with the known results for $\W_N$; this is sketched below in 
section~\ref{sec:minreps}.
Actually, effectively the same analysis was already done in 
\cite{Hornfeck:1992dp,Hornfeck:1993kp} (and later in \cite{Blumenhagen:1994wg}), 
leading to 
\be \label{c433}
(C^4_{33})^2 \equiv \gamma^2 = 
\frac{64 (c+2) (\mu-3) \bigl( c(\mu+3) + 2 (4\mu+3)(\mu-1) \bigr)}{
(5c+22) (\mu-2)  \bigl( c(\mu+2) + (3\mu+2)(\mu-1)\bigr)} \ .
\ee
Note that there is a sign ambiguity in the definition of $C^4_{33}$ since the 
normalisation convention of \cite{Hornfeck:1992tm} is defined by 
fixing the OPE of the spin $s$ field $W^{(s)}$ with itself
\be
W^{(s)} \cdot W^{(s)} \sim \frac{c}{s}\cdot  {\bf 1} + \cdots \ ,
\ee
and hence only determines the normalisation of each field up to a sign. The same comment
also applies to the other structure constants (see below).

In \cite{Hornfeck:1992dp,Hornfeck:1993kp,Blumenhagen:1994wg}
a few of the other low-lying structure 
constants of $\w{\mu}$ were also derived; in the conventions of \cite{Hornfeck:1992tm} 
and  using our notation they are explicitly equal to (see eqs.\ (2.1.25a/b) of 
\cite{Blumenhagen:1994wg} as well as  \cite{Hornfeck:1994is})
\begin{eqnarray}\label{othstr}
C^4_{33} C^4_{44} & = & 
\frac{48\bigl( c^2 (\mu^2-19) + 3 c (6 \mu^3 - 25 \mu^2 + 15)
+ 2 (\mu-1)(6\mu^2 - 41\mu-41) \bigr)}{ (\mu-2) (5c+22) \,
\bigl( c (\mu+2) + (3\mu+2)(\mu-1) \bigr)}\hspace*{0.9cm}  \label{c444} \\[4pt]
(C_{34}^5)^2 & = & 
\frac{25 (5c+22) (\mu-4) \bigl( c (\mu+4) + 3 (5\mu+4)(\mu-1) \bigr)}{(7c+114)
(\mu-2) \, \bigl( c(\mu+2) + (3\mu+2)(\mu-1)\bigr)} \\[4pt]
C_{45}^5 & = & \frac{15}{8 (\mu-3) (c+2) (114 + 7c)\bigl(c(\mu+3)
+ 2 (4\mu+3) (\mu-1)\bigr)} \, C^4_{33} \,
 \nonumber \\ 
& & \times  \Bigl[ c^3 (3 \mu^2-97)
+ c^2(94 \mu^3 - 467 \mu^2 -483)
+ c (856 \mu^3 - 5192 \mu^2 + 4120) \nonumber \\
& & \quad 
+ 216 \mu^3 - 6972 \mu^2 + 6756\Bigr] \ . \label{c455}
\end{eqnarray}
These expressions look very complicated, but as we will see momentarily,
they actually exhibit a very nice structure.

\subsection{Triality in $\w{\mu}$}

Our first observation is that, for fixed $c$, there are three values of 
$\mu$ (which we label as $\mu_{1,2,3}$) for which the structure constant 
$\gamma$ in (\ref{c433}) is actually the same. Indeed, for given $c$ and $\gamma$, it
follows directly from (\ref{c433}) that the three values  are 
the roots of the cubic equation 
\be
\label{lambcub}
\bigl(3\tilde{\gamma}^2-8\bigr)\, \mu^3
+\bigl(\tilde{\gamma}^2(c-7)+(26-c)\bigr)\, \mu^2
-\bigl( 4\tilde{\gamma}^2(c-1)-9(c-2)\bigr)=0 \ ,
\ee
where we have defined $\tilde{\gamma}^2=\gamma^2\frac{(5c+22)}{64(c+2)}$.
Note that the cubic equation does not have a linear term in $\mu$; thus the three
solutions satisfy 
\be
\mu_1 \mu_2 + \mu_2 \mu_3 + \mu_3 \mu_1 = 0 \ ,
\ee
which is equivalent to $\sum_{i=1}^{3} \tfrac{1}{\mu_i}=0$ provided that
all $\mu_j\neq 0$. 

The analysis from the beginning of this section suggests that actually the full
$\w{\mu}$ algebra should exhibit this triality symmetry, i.e.\ that 
{\em all} structure constants are the same for $\mu_{1,2,3}$. At least for the known
structure constants in eqs.~(\ref{c444}) -- (\ref{c455}) this is true; one way to 
see this, is to observe that they can all be expressed in terms of $\gamma$ and 
$c$ as 
\begin{eqnarray}
C^4_{44} & = &  \frac{9(c+3)}{4(c+2)}\, \gamma - \frac{96 (c+10)}{(5c+22)}\, \gamma^{-1} \\[4pt]
(C_{34}^5)^2 & = & \frac{75 (c+7) (5c+22)}{16 (c+2) (7c+114)}\, \gamma^2 -25 \\[4pt]
C_{45}^5 & = &  \frac{15\, (17 c + 126) (c + 7)}{8\, (7 c + 114) (c + 2) } \, \gamma
-240 \frac{(c+10)}{(5c+22)}\, \gamma^{-1} \ .
\end{eqnarray}
Incidentally, the structure of these identities suggests that these higher OPE coefficients
are completely determined from $C^4_{33}$ by the  Jacobi identities, and this appears 
indeed to be true \cite{CGKV}. 
Since the three values of $\mu_{1,2,3}$ lead to the same value of $\gamma$
(at a given $c$), these structure constants are then also equal for the three values of 
$\mu$.  This is a very strong indication that the quantum $\w{\mu}$ algebras are 
actually equivalent for these three (generically distinct) values of $\mu$, i.e.\ that 
\be\label{keyrelation}
\boxed{
\w{\mu_1} \cong \w{\mu_2} \cong \w{\mu_3} \qquad \hbox{at fixed $c$}}
\ee
where $\mu_{1,2,3}$ are the roots of the cubic equation \eqref{lambcub}, evaluated
for a given $\gamma$.

Note that these algebras look very different from the point of view of $\hs{\mu}$ 
or even at the classical level. In fact, at very large $c$, eq.~(\ref{lambcub}) reduces to a 
linear equation in  $\mu^2$, and hence reduces to the familiar equivalence
between the classical $\w{\mu}$ algebras for $\pm \mu$ ---  this 
property is directly inherited from $\hs{\mu}$. The statement in (\ref{keyrelation}) is 
a very nontrivial generalisation to the quantum level (finite $c$), where 
the equivalence is a triality between the three values $\mu_{1,2,3}$. There
are three special cases where the cubic equation eq.~(\ref{lambcub}) 
degenerates: for $\mu=0$ we have 
$\tilde\gamma^2 = \tfrac{9 (c-2)}{4(c-1)}$, 
and the constant term in (\ref{lambcub}) vanishes. Then $\mu=0$ is a double
zero, and the other solution simply becomes 
\be
\w{\mu=0} \ \cong \ \w{\mu = c+1} \ .
\ee
For $\mu=1$, on the other hand, we have $\tilde\gamma^2 = \tfrac{8}{3}$, and the
cubic power vanishes; then we have the equivalences
\be
\w{\mu=1} \ \cong \ \w{\mu = -1}  \ \cong \ \w{\mu=\infty} \ .
\ee
The fact that for $\mu=1$ the symmetry $\mu\mapsto -\mu$ survives
at the quantum level is a direct consequence of the fact that, for this value of 
$\mu$, $\w{\mu}$ is a linear $\W$-algebra whose structure constants 
are simply the (analytic continuation) of the $\hs{\mu}$ structure constants.

Finally, the coefficient in front of the $\mu^2$ term in (\ref{lambcub}) vanishes
for $\tilde\gamma^2 = \tfrac{(c-26)}{(c-7)}$, when the equation becomes
$\mu^3=(c+1)$. Thus the three cubic roots of $(c+1)$ define equivalent 
$\w{\mu}$ algebras.

\subsection{Truncation to Finite $N$}

In order to clarify the analytic continuation of \cite{Castro:2011iw} in section~\ref{sec:analytic}
we will be interested in the case where the algebra $\w{\mu}$ truncates to $\W_N$. In
that case, the coset level-rank duality from the beginning of this section 
suggests that we have the equivalences (\ref{tria1}). 
We now want to show that they are a special case of (\ref{keyrelation}). 

In order to see this we take one of the roots of 
(\ref{lambcub}) to be $\mu_1=N$. Then this determines 
$\gamma =\gamma(\mu=N, c)$, and hence the other two roots
$\mu_{2,3}$. It follows from (\ref{lambcub}) that they satisfy 
the quadratic equation
\be\label{muceq}
\mu^2(N^2-1)-\mu(N-1-c)- N(N-1-c) =0\ ,
\ee
whose solutions are 
\be
\mu_{2,3}(N, c) = \frac{1}{2(N^2-1)} \Bigl[
(N-1-c) \pm \sqrt{(N-1-c)(4 N^3 - 3N - c -1)} \Bigr] \ .
\ee
For the particular value $c=c_{N,k}$ defined in (\ref{cNk}), we then find
\be\label{Nkequiv}
\mu_2(N, c_{N,k})= \frac{N}{N+k} \qquad\hbox{and}
\qquad \mu_3(N, c_{N,k})= -\frac{N}{N+k+1}  \ , 
\ee
thus reproducing precisely (\ref{tria1}). In particular, this therefore gives strong evidence for the
equivalence of the ${\cal W}_{N}$ minimal model at level $k$,
with the $\w{\mu}$ theory at $\mu=\lambda\equiv \tfrac{N}{N+k}$. This symmetry is crucial
for the duality to the bulk Vasiliev theory proposed in \cite{Gaberdiel:2010pz}.

Moreover, we see that there is another value of $\mu$, namely 
$\mu_3= -\tfrac{N}{N+k+1}$, which is also equivalent to the other two descriptions. In the large 
$N$ 't~Hooft limit, $\mu_2=-\mu_3$ and this is just the statement about the classical equivalence of the $\hs{\pm \mu}$ theories.

\section{Minimal Representations of $\w{\mu}$}\label{sec:reps}

In this section, we will study a special class of representations of $\w{\mu}$ 
and see how the results are consistent with the above equivalences. 
These considerations will also play an important role for
the analysis of the analytic continuation in section~\ref{sec:analytic}.

\subsection{Determining the $c$-Dependence of the Structure Constants}\label{sec:minreps}

Let us consider the representation of $\w{\mu}$ that has the fewest number
of low-lying states. Leaving aside the vacuum representation, this `minimal'
representation has then the character 
\be\label{minchar}
\chi = \frac{q^h}{(1-q)} \prod_{s=2}^{\infty} \prod_{n=s}^{\infty} \frac{1}{(1-q^n)} 
= q^h \Bigl( 1 + q + 2 q^2 + \cdots \Bigr)\ . 
\ee
Note that if the conformal dimension of the ground state is non-zero, then the $L_{-1}$
descendant is necessarily non-trivial, and hence the representation contains at least
one state at level one; for the `minimal' representation this is the only non-trivial
state at level one, i.e.\ all other states are proportional to it (modulo null states). Similarly,
in the minimal representation there are only two descendants at level two, which we 
may take to be the  $L_{-1}^2$ and $L_{-2}$ descendants of the ground state; all
other descendants at level two are again equal to linear combinations of them (modulo
null states). 

Thus the minimal representation must contain many (sic!)~null-vectors, and as
a consequence its structure is completely determined. In particular, 
it follows from the analysis of appendix~\ref{app:rep} that the conformal dimension
of the ground state must satisfy the cubic equation
\begin{eqnarray}\label{hcube}
& \hspace*{-4.5cm} 
7(5c+22)(16h^2+2ch+c-10h)(2 c h - 3 c - 2 h) N_4 \\
& \qquad 
- 150(18 h^2 c + c^2 - 12c h  + c^2 h  +36h^2+2c-28h)
(hc-2h-2c)N_3^2 =0 \ . \nonumber 
\end{eqnarray}
We can then turn the logic around, and {\em use} this identity to 
determine the full $c$-dependence of
the structure constants $N_3$ and $N_4$.\footnote{Actually, as is 
clear from the structure of the $\w{\mu}$ algebra, there is an overall normalisation
freedom (which is described by $q^2$ in (\ref{N3c}) and (\ref{N4c}) and which
corresponds to rescaling the primary field of spin $s$ by $q^{s-2}$), and
only $N_4/N_3^2$ has any independent meaning.} To this end we recall that
$\w{\mu=N}$ truncates to $\W_N$, and that the $\W_{N,k}$ theories have
the minimal representations $({\rm f};0)$ and $(0;{\rm f})$ (or
their conjugates), where ${\rm f}$ denotes the fundamental representation of 
$\mathfrak{su}(N)$. The corresponding conformal dimensions equal
\be\label{hf}
h({\rm f};0) = \frac{N-1}{2N} \Bigl( 1 + \frac{N+1}{N+k} \Bigr) \ , \qquad
h(0;{\rm f}) = \frac{N-1}{2N} \Bigl( 1 - \frac{N+1}{N+k+1} \Bigr) \ . 
\ee
Either of these values must therefore be a solution of 
(\ref{hcube}) for {\rm finite} $N$ and $k$; expressing $k$ in terms of $c$ and 
$N=\mu$, the first value of $h$ in (\ref{hf}) implies that we have 
\be\label{Nreln}
\frac{N_4}{N_3^2} =    
\frac{75\, (c+2)\, (\mu-3)\, \bigl(c(\mu+3)+2(4\mu+3)(\mu-1)\bigr) }
{14\, (5c+22) \, (\mu-2)\, \bigl(c(\mu+2)+(3\mu+2)(\mu-1)\bigr)} \ . 
\ee
It is then a non-trivial consistency check that, with this expression for the structure
constants, also the second value of (\ref{hf}) solves
(\ref{hcube}). In order to relate this to the conventions of \cite{Hornfeck:1992tm}, 
we choose $N_3=\tfrac{2}{5}$ so that the OPE of two $W$-fields is as in
(\ref{WWOPE}), and rescale $U$ as in (\ref{betadef}); then (\ref{Nreln}) implies that 
\be \label{c433a}
(C^4_{33})^2 =   \frac{16 \cdot 56}{75}\, \frac{N_4}{N_3^2} = 
\frac{64 (c+2) (\mu-3) \bigl( c(\mu+3) + 2 (4\mu+3)(\mu-1) \bigr)}{
(5c+22) (\mu-2)  \bigl( c(\mu+2) + (3\mu+2)(\mu-1)\bigr)} \ ,
\ee
thus leading to (\ref{c433}).

\subsection{Structure of Solutions}

Plugging in the explicit expressions for $\tfrac{N_4}{N_3^2}$, the cubic equation in 
(\ref{hcube}) factorises into a linear equation
\be\label{hlin}
2h\, (1-\mu+c)-(1+\mu)c=0\qquad \Longrightarrow  \qquad 
h = h^{(1)}(\mu,c) \equiv \frac{(1+\mu)c}{2(1+c-\mu)} \ , 
\ee
as well as the quadratic equation
\be\label{hquad}
4h^2\mu^2+2h(1+c+\mu-2\mu^2)-c(1-\mu)=0
\ee
with solutions
\be\label{h2pm}
h = h^{(2)}_{\pm} (\mu,c)\equiv 
 \frac{1}{4\mu^2} \Bigl[
- (1+c+\mu-2\mu^2) \pm \sqrt{(c+1- \mu)\, (c+1+3\mu-4\mu^3)}\Bigr] \ .
\ee
Note that the cubic equation (\ref{hcube}), once we substitute (\ref{Nreln}), is
actually triality invariant. As a consequence the roots in (\ref{hquad}) and (\ref{h2pm}) are 
permuted among each other under a triality transformation.
 
For $c\rightarrow \infty$ (and $\mu$ fixed), the three solutions behave as 
\be\label{h3rd}
h^{(1)} \simeq \frac{1}{2} (1+\mu) \ , \qquad
h^{(2)}_{+} \simeq \frac{1}{2} (1-\mu) \ ,  \qquad
h^{(2)}_{-}\approx -\frac{c}{2\mu^2}  + \frac{\mu^3+\mu^2-\mu-1}{2\mu^2}\ . 
\ee
For $\mu=\lambda\equiv \frac{N}{N+k}$, the first two are the familiar solutions 
for the scalar fields in the duality of 
\cite{Gaberdiel:2010pz}, while the last solution does not correspond to a representation 
of $\hs{\lambda}$. The reason for this is that, as discussed in detail in 
\cite{Gaberdiel:2011wb}, $\hs{\lambda}$ is only a subalgebra of $\w{\lambda}$ for 
$c\rightarrow \infty$, but the third representation in (\ref{h3rd}) decouples in this limit. 

We will see in the next subsection that for $\mu=N$ and $c$ taking one of the values 
$c_{N,k}$ of the $\W_{N,k}$ minimal models, $h^{(2)}_{\pm}$ correspond to the physical 
representations $({\rm f};0)$ and $(0;{\rm f})$, respectively. Note the different behavior 
of the dimensions of the two representations at large central charge, suggesting
that the two scalar excitations appear on a very 
different footing. We will return to this important distinction in section~5.

\subsection{The Minimal Model Parametrisation}\label{finiteNsub}

For the following it will sometimes be useful to parametrise the $\w{\mu}$ 
algebra in terms of $N$ and $k$, rather than $c$ and $\mu$, where the relation
between the two parameters is that $\mu=N$, and $c=c_{N,k}$, with 
$c_{N,k}$ defined in (\ref{cNk}). Obviously, this is a useful parametrisation if we 
are interested in the truncation of $\w{\mu}$ to finite $N$ (as will be the case in
section~\ref{sec:analytic}). However, we may also take the more general point of 
view that $N$ and $k$ are not necessarily integers, and then this is just a useful
parametrisation (covering all $\w{\mu}$ algebras at all values of the central charge). 

We should note, however, that this parametrisation exhibits a six-fold ambiguity. 
The reason for this is that, for a given $c$ and $N$, the equation $c=c_{N,k}$ 
has two solutions for $k$; if  $k=k'$ is one solution, then
the second solution is $k= - (2N+k'+1)$. Together with (\ref{Nkequiv}), the 
six equivalent pairs are therefore
\be
\begin{array}{lll}
(N,k) & \qquad & (N,-2N-1-k) \\[4pt]
{\displaystyle \bigl( \tfrac{N}{N+k},\tfrac{1-N}{N+k} \bigr)} & \qquad &   
{\displaystyle \bigl( \tfrac{N}{N+k},- \tfrac{2N+k+1}{N+k} \bigr)} \\[8pt]
{\displaystyle \bigl( - \tfrac{N}{N+k+1},- \tfrac{k}{N+k+1} \bigr)} & \qquad & 
{\displaystyle \bigl( - \tfrac{N}{N+k+1}, \tfrac{N-1}{N+k+1} \bigr) \ .}
\end{array}
\ee  
In this parametrisation the eigenvalues of the minimal representations are 
\begin{eqnarray}
h({\rm f};0) & = &  \frac{N-1}{2N} \Bigl( 1 + \frac{N+1}{N+k} \Bigr) = 
h^{(2)}_+(\mu=N,c=c_{N,k})  \label{minmodiden1} \\
h(0;{\rm f}) & = &  \frac{N-1}{2N} \Bigl( 1 - \frac{N+1}{N+k+1} \Bigr) = 
h^{(2)}_-(\mu=N,c=c_{N,k}) \ ,  \label{minmodiden2}
\end{eqnarray}
while the third solution, eq.~(\ref{hlin}), equals
\be\label{thirdsol}
h^{(1)}(\mu=N,c=c_{N,k}) =  - k\, \frac{ (2N+k+1)}{2N} \ .
\ee
This last representation does not appear to be (and in fact is not) a representation
of $\W_N$ (at level $k$). It may therefore seem that we have a contradiction 
with (\ref{tria1}). 

In order to understand why this is not the case, we need to be more
precise about the nature of the truncation, say in (\ref{obvious}). This
identity is only true after quotienting $\w{\mu}$ by 
the non-trivial ideal that appears for $\mu=N$; this just mirrors the fact 
that $\hs{\mu=N}$ is not identically equal to 
$\mathfrak{su}(N)$ either since $\hs{\mu}$ is infinite dimensional, whereas 
$\mathfrak{su}(N)$ is finite-dimensional. Rather, for $\mu=N$ the algebra $\hs{\mu}$ develops 
an (infinite-dimensional) ideal, and if we divide $\hs{\mu}$ by this ideal, the resulting Lie 
algebra is isomorphic to $\mathfrak{su}(N)$.

Similarly, for (\ref{obvious}), (\ref{claim}) or (\ref{tria1}), the two algebras are
only isomorphic if we quotient $\w{\mu}$ by the relevant ideal (that appears for these
special values of $\mu$). But then it is not guaranteed that the representations of 
$\w{\mu}$ are compatible with this quotienting. 

For example, if we set $\mu=3$, then $\w{3}$ should truncate to ${\cal W}_3$. 
This requires that we set all higher spin fields with spin greater than $3$ 
(such as $U$ and $X$) to zero.  But then only those `minimal'
representations of $\w{\mu}$ define (minimal) representations of ${\cal W}_3$ for which
$u=x=0$, see (\ref{uh}) and (\ref{reln1}). In particular, the numerator of $u$ in 
(\ref{uh}) leads to a quadratic relation for $h$, that is satisfied for $h=h^{(2)}_{\pm}$,
but not for $h=h^{(1)}$. [Similarly, (\ref{reln1}) then also follows since $N_4=0$
for $\mu=3$, see eq.\ (\ref{Nreln}).] This explains why (\ref{thirdsol}) is {\em not} a
representation of $\W_N$ at $N=3$, and we expect that a similar argument 
will apply for any integer $N$. Similarly, if we set 
\be
\mu=\frac{3}{3+k} \qquad \hbox{with}\qquad c=c_{3,k} = 2 - \frac{24}{(3+k)(4+k)} \ ,
\ee
corresponding to ${\cal W}_3$ at  level $k$, then $h^{(1)}$ and $h^{(2)}_{+}$ 
describe the actual representations of  ${\cal W}_3$ at that central charge, 
whereas $h^{(2)}_{-}$ does not, as can be seen by the same
reasoning. Note that also for these values of $\mu$ and $c$, $N_4$ in 
(\ref{Nreln}) vanishes.

\section{Analytic Continuation}\label{sec:analytic}

Let us now apply the insights of the previous sections 
to shed some more light on the analytic continuation proposed in
\cite{Castro:2011iw}. Recall that this analytic continuation related a class 
of states in the ${\cal W}_{N,k}$ minimal models (the `light states') to certain classical 
solutions in the (euclidean) higher spin theory based on the gauge group 
${\rm SL}(N, \mathbb{C})$. In the process $N$ was kept fixed and finite, while
$k$, which is a positive integer in the minimal models, was taken to the value $k=-(N+1)$. 
The  expressions for the dimensions as well as spin $3$ and spin $4$ charges of all these 
states in the minimal model were formally found to match (in a fairly non-trivial way) 
with those of the bulk solutions for any value of $N$. 

It is not immediately obvious whether the formal procedure of taking $k$ from positive integer 
values to the negative value $k=-(N+1)$ makes any sense, and indeed, in the analysis of
\cite{Castro:2011iw}, it was not entirely clear what precisely was being kept fixed in
the process. With our
improved understanding of the structure of the quantum $\w{\mu}$ algebras we can now give
a clear interpretation of these results. 
As we shall explain below, the correct way to describe this analytic continuation is to
consider the family of $\W_N$ theories at fixed finite $N$, and vary $c$ from the
minimal model value $c\leq (N-1)$ to the semiclassical case where $c\rightarrow \infty$. 
Since we now understand how to describe the algebras $\w{\mu}$ for arbitrary 
$\mu$ and $c$, this analytic continuation is well-defined, and it induces a corresponding
analytic continuation on all representations.

In the following we shall first study this in detail for the lightest of the light states, the
representation labelled by $(\f,\f)$. This then suggests a natural generalisation for
all the light states; this will be described in section~\ref{sec:anagen}. 
Incidentally, our analysis also implies that the two minimal
representations (that play an important role in the duality of \cite{Gaberdiel:2010pz})
behave rather differently as we take $c\rightarrow \infty$. This suggests that 
one of them should probably not be interpreted as a `perturbative' state of the 
higher spin gravity theory; we will come back to this issue in section~\ref{sec:concl}.

\subsection{The $(\f,\f)$ States}\label{sec:ff}

Recall that the primaries of the minimal models are labelled by two integrable 
representations $(\Lambda_+; \Lambda_-)$ of $\mathfrak{su}(N)_k$ and 
$\mathfrak{su}(N)_{k+1}$, respectively. The set of light states that were considered in 
\cite{Castro:2011iw} are of the form $\Lambda_+=\Lambda_-=\Lambda$, and their
conformal dimension equals 
\be
h(\Lambda, \Lambda)=\frac{c_2(\Lambda)}{(N+k)(N+k+1)} \ .
\ee
Here $c_2(\Lambda)$ is the quadratic Casimir of the representation $\Lambda$.
If we were to consider the 't~Hooft limit where $(N,k) \rightarrow \infty$, 
the states where $\Lambda$ has a finite number of Young tableau boxes have
conformal dimension $h \propto \tfrac{1}{N}$, and hence were dubbed `light states'.
In the following we want to follow these states as we change the conformal dimension
from the minimal model value $c=c_{N,k}$ to the quasiclassical regime where $c$ is large. 
Let us first explain this in detail for the lightest of the light states, the one corresponding
to $\Lambda=\f$. 

In order to do so we note that we can think of $(\f;\f)$ as the fusion of $(0;\f)\otimes (\f;0)$. 
We can thus repeat the fusion analysis of \cite{Gaberdiel:2011zw}, but now
done for $\w{\mu=N}$ at finite $c$, using the explicit form of the commutation
relations of appendix~\ref{app:comm}; some details of this calculation are given in
appendix~\ref{app:fus}. Provided that $c$ is finite, the 
resulting fusion product is irreducible,\footnote{This is different to what happened in  
\cite{Gaberdiel:2011zw} where the $c\rightarrow \infty$ limit was considered. There 
the fusion product turned out to be indecomposable.} and the conformal dimension
of the resulting highest weight state equals exactly 
\be\label{fusion}
h(\f;\f) = h(\f;0) + h(0;\f) - \frac{N-1}{N} \ .
\ee
Given the identifications (\ref{minmodiden1}) and (\ref{minmodiden2}), 
we know how to analytically continue both $(\f;0)$ and $(0;\f)$; for
$c\rightarrow \infty$, it then follows that their conformal dimensions 
behave as (see eq.~(\ref{h3rd}))
\be\label{minlim}
h({\rm f};0) \sim - \frac{(N-1)}{2} \ , \qquad
h(0;{\rm f}) \sim - \frac{c}{2 N^2} + \frac{N^3+N^2-N-1}{2N^2} 
\ , \qquad \hbox{as $c\rightarrow \infty$.}
\ee
Thus it follows that the analytic continuation of the $(\f;\f)$ representation has
conformal dimension
\be\label{ffans}
h(\f;\f) \sim - \frac{c}{2N^2} + \frac{N-1}{2N^2} \qquad \hbox{as $c\rightarrow \infty$.}
\ee
This then reproduces precisely the observation of \cite{Castro:2011iw}.

\subsection{Light States and Conical Surpluses}\label{sec:anagen}

One can actually generalise the above discussion to all light states. To this end one 
observes that the conformal dimension, as well as the eigenvalues
of the spin $3$  zero mode of the state $(\Lambda;\Lambda)$,
can be written as 
\begin{eqnarray}
h & = &  \alpha_0^2\, c_2(\Lambda)  \nonumber \\
w &=& \alpha_0^3 \, c_3(\Lambda+\rho) \ , \label{eigenvals}
\end{eqnarray}
where $\rho$ denotes the Weyl vector of the algebra $\mathfrak{su}(N)$, $c_s$ are
the various Casimir operators --- for the precise definitions see eqs.~(4.12) 
and (5.10) of \cite{Castro:2011iw} --- and $\alpha_0$ is defined by 
\be\label{alpha0Nk}
\alpha_0^2=\frac{1}{(N+k)(N+k+1)} \ . 
\ee
Note that the entire $k$ dependence of the expressions in (\ref{eigenvals}) is contained 
in the prefactor $\alpha_0$. We should mention in passing that in  \cite{Castro:2011iw}
a similar statement was also made (to leading order in $\tfrac{1}{c}$) for the spin $4$ zero mode.
This (as well as corresponding statements for the higher spin charges) 
can be deduced from the Drinfeld-Sokolov description, using
the simple formula for the eigenvalue in the non-primary basis, see for example
eq.\  (6.50) of \cite{BouScho93}. However, the field redefinition that is required 
for going from this non-primary basis to the corresponding primary basis 
is only known to leading order in $\tfrac{1}{c}$ \cite{DiFrancesco:1990qr}, and is likely
to receive non-trivial quantum corrections (coming from the normal ordering) for higher
spins.  Thus the simple statement corresponding to (\ref{eigenvals}) will, for
spins greater than three, only hold in a suitable non-primary basis.

Returning to (\ref{eigenvals}), it is now 
natural to believe that the analytic continuation simply consists of writing these
expressions in terms of $N$ and $c$, rather than $N$ and $k$. Since 
\be
c= c_{N,k}= (N-1) \, \Bigl[ 1 - \frac{N (N+1)}{(N+k) (N+k+1)} \Bigr] 
=(N-1) \, \Bigl(1 - N (N+1)\, \alpha_0^2\Bigr) \ ,
\ee 
this amounts to writing 
\be\label{alpha0def}
\alpha_0^2 =  \frac{(N-1-c)}{N(N^2-1)} \ .
\ee
The analytic continuation is then straightforward: we keep $N$ fixed, and vary $c$
continuously from $c=c_{N,k}$ to the semiclassical regime $c \rightarrow \infty$. Note
that for the case of $\Lambda=\f$ this then reproduces indeed (\ref{ffans}) since
$c_2(\f) = \tfrac{N^2-1}{2N}$. Similarly, it follows from the analysis of 
appendix~\ref{app:fus} that the same holds for $w(\f;\f)$, see eq.\ (\ref{wfffin}).

Since the entire $c$ dependence of the eigenvalues (\ref{eigenvals})
is carried by their  dependence on $\alpha_0$, and since $\alpha_0^2 \sim c$
for large $c$, all the   eigenvalues in (\ref{eigenvals}) become proportional
to some positive power of $c$. Thus these states can be interpreted in terms of 
`classical solutions' in this limit.

More concretely, the $c$ dependence of $\alpha_0$ 
(at fixed $N$) contains only a linear and constant term
\be
\label{alphc}
\alpha_0^2(c)  = -\frac{c}{N(N^2-1)} + \frac{1}{N(N+1)} \ .
\ee
In the semiclassical limit we can drop the second (constant term), and 
with the resulting value of $\alpha_0^2$, the spectrum and charges were matched with that of 
the conical defect solutions in the bulk higher spin theory. This match is only true to 
leading order in the central charge;\footnote{We thank Joris Raeymakers for sharing 
closely related observations.} of course, that is the best one could hope for from the 
classical solutions which are not sensitive to $\frac{1}{c}$ corrections. 
In fact, our analysis gives a prediction that the energies of the 
conical defects only get an ${\cal O}(1)$ positive correction
\be
\delta h(\Lambda, \Lambda) = \frac{c_2(\Lambda)}{N(N+1)} \ ,
\ee
without any further $\frac{1}{c}$ corrections. However, the higher spin charges given 
in (\ref{eigenvals}) do generically have higher order corrections. But perhaps all these 
corrections are best viewed as a quantum renormalisation of the central charge 
$c \rightarrow c-(N-1)$ (or equivalently Newton's constant $G_N$)  in the bulk theory.  

\subsection{Analytic Continuation of the Minimal States}\label{sec:anaper}

Actually, the above considerations also apply directly to the `minimal'
representations; indeed, this is already implicit in what was done above in 
section~\ref{sec:ff}. As we explained there, the conformal dimension of the 
two minimal representations that exist at finite integer $N$ behave as 
(\ref{minlim}). Note that the semiclassical interpretation of these two states (at fixed $N$)
is quite different: the conformal dimension of the state $({\rm f};0)$ remains 
finite, while that of $(0;{\rm f})$  is proportional to $c$. 
We are therefore led to the point of view that the only true perturbative states are 
those corresponding to $({\rm f};0)$ and $(\bar{\rm f}; 0)$, making up one complex scalar. 
What was earlier interpreted as another perturbative scalar, namely, that corresponding
to  $(0;{\rm f})$ and its conjugate $(0;\bar{\rm f})$, is perhaps more naturally thought 
of as a solitonic state 
that just happens to have a finite dimension in the 't~Hooft limit.  This is also in line with the observations made at the end of section~3.2. We shall come back to 
the implications of this for the duality proposed in
\cite{Gaberdiel:2010pz}  in section~\ref{sec:concl}.

Incidentally, eq.~(\ref{fusion}) also leads to a somewhat different point of view. Since
in the large $c$ limit $h(\f;0) = h(\bar{\f};0) <0$, it is more natural to rewrite 
(\ref{fusion})  as 
\be
h(0;\f) = h(\f;\f) - h(\bar{\f};0) +  \frac{N-1}{N}  \ .
\ee
This suggests that $(0;\f)$ should be interpreted as some kind of bound state of 
$(\f;\f)$ with a perturbative excitation $(\bar{\f};0)$. Actually, a similar statement
holds for all representations of the form $(0;\Lambda_-)$. To see this, recall that
the conformal dimension of the representation $(\Lambda_+;\Lambda_-)$ equals
\be
h(\Lambda_+;\Lambda_-) = \frac{1}{2} (\Lambda,\Lambda+2 \alpha_0 \rho) \ ,
\qquad \hbox{where} \qquad
\Lambda = \alpha_+ \Lambda_+ + \alpha_- \Lambda_-  \ ,
\ee
the inner product $(\cdot,\cdot)$ is the usual inner product on the weight space, 
and
\be
\alpha_+ = \frac{1}{\sqrt{k_{\rm DS}+N}} \ , \qquad
\alpha_- = - \sqrt{k_{\rm DS}+N} \ , \qquad \alpha_0 = \alpha_+ + \alpha_- \ . 
\ee
Note that, using the relation to the coset labels of eq.~(\ref{DSrel}), $\alpha_0^2$ agrees
then precisely with (\ref{alpha0Nk}). Since $\alpha_+ \cdot \alpha_- = -1$, it now
follows that
\be\label{0lam}
h(\Lambda_+;\Lambda_-) = h(\Lambda_+;0) + h(0;\Lambda_-) - (\Lambda_+,\Lambda_-) \ .
\ee
In particular, we can apply this to the case $\Lambda_+=\Lambda_-$ and conclude that 
\be\label{Lamm}
h(0;\Lambda_-) = h(\Lambda_-;\Lambda_-) - h(\Lambda_-^\ast;0) + (\Lambda_-,\Lambda_-) \ .
\ee
The last term is positive and purely group-theoretic, i.e.\ it does not depend on $k$ (or $c$),
but only on $N$ (as well as $\Lambda_-$). We can thus think of 
$(0;\Lambda_-)$ as a bound state of $(\Lambda_-;\Lambda_-)$ with the perturbative
excitation $(\Lambda_-^\ast;0)$. Note that, again, $h(\Lambda;0)$ becomes
negative in the semiclassical limit, since we have 
\be
h(\Lambda;0) = \alpha_+^2 \, c_2(\Lambda) - (\Lambda,\rho) \ , 
\ee
where 
\be
\alpha_+^2 = \frac{k+N+1}{k+N} \ \rightarrow \ 0 \qquad
\hbox{as $k\rightarrow - N -1$.}
\ee

\section{Refining the minimal model holography conjecture}\label{sec:concl}

One of the striking consequences of the analysis of the 
quantum $\w{\mu}$ algebra is the very different nature of the two minimal 
representations which correspond to $(\f; 0)$ and $(0;\f)$, respectively.
This difference was not at all obvious in the 't~Hooft limit where they have dimensions 
$\half(1\pm \lambda)$, and appear to be on a similar footing. However, we 
now see that this is, in a sense, an artifact of the 't~Hooft limit: for 
any finite $N$ (i.e.\ $\mu=N$) the two states have conformal dimensions with a very 
different dependence on $c$; in particular, in the semi-classical limit they behave as 
\be\label{minlim1}
h({\rm f};0) \sim \frac{(1-N)}{2} \ , \qquad
h(0;{\rm f}) \sim - \frac{c}{2 N^2} \ , \qquad \hbox{as $c\rightarrow \infty$} \ ,
\ee
see eq.\ (\ref{minlim}). Since $h(0;\f) \propto c$, this now suggests that 
$(0;\f)$ is more naturally thought of as a non-perturbative state or soliton, rather
than a perturbative state. We should mention that both $h(\f;0)$ and $h(0;\f)$ 
turn negative in this limit (i.e.\ for  $c\rightarrow \infty$ at fixed $N$), 
thus signalling that the theory becomes non-unitary. However, for the purpose 
of identifying the semiclassical interpretation of the various states, this should 
be immaterial. (Obviously,  this problem is absent in the actual 't~Hooft limit 
since it is a sequence of unitary minimal models; it corresponds to taking
both $N$ and $c$ to infinity in a 't~Hooft like manner.)

It is therefore natural to propose that 
the state $(0;\f)$ (and its conjugate) should {\em not} be thought of as  corresponding 
to a perturbative scalar mode in the bulk. Instead, there is only one (complex) perturbative 
scalar dual to $(\f; 0)$ (and its conjugate). We should then think of the state $(0;\f)$ as 
being on the same footing as the light states such as $(\f; \f)$. Indeed, as we saw in the 
previous section, in the large $c$ (finite $N$) limit it makes a lot of sense to view $(0;\f)$ as a 
bound state of $(\f; \f)$ with the perturbative state $(\bar{\f};0)$.
Since $(\f; \f)$, was already identified with a semiclassical solution (a conical surplus), we 
then also have a candidate bulk interpretation for the state $(0;\f)$ as an excitation of this 
semi-classical solution. 

This reinterpretation also makes sense of the observations in 
\cite{Papadodimas:2011pf, Chang:2011vk},
where it was argued that the double trace operator corresponding to the two particle 
state of $(0;\f)$ and $(\f; 0)$ 
(whose conformal dimension equals $h=1$ in the 't~Hooft limit) is a descendant of the 
light state $(\f; \f)$. This would be very strange if one were to interpret both $(0;\f)$ and $(\f; 0)$ as 
perturbative states while viewing $(\f; \f)$ as non-perturbative. The interpretation
we are proposing here, on the other hand, makes this quite natural since we now 
consider $(\f; \f)$ as the basic non-perturbative object, which has $(0;\f)$ as an
excited state.

Finally, this also fits in with the fact that the most natural (unambiguous)  
bulk higher spin $\hs{\lambda}$ theory is the one with a single complex scalar. 
This was one of the motivations for the proposal of \cite{Chang:2011mz} that 
this higher spin theory describes, say, the $(\Lambda; 0)$\footnote{In their analysis it was 
ambiguous whether to consider the $(\Lambda; 0)$ or $(0; \Lambda)$ subsector of the 
minimal models, whereas we see here a basic distinction between the two.} sector of the 
minimal models. In a similar vein, at $\lambda=0$ it is the theory with a single complex
scalar that is dual to the singlet sector of a free theory \cite{Gaberdiel:2011aa}.

The final picture is therefore one in which the bulk theory has a perturbative sector 
consisting of one complex scalar (dual to $(\f; 0)$ and its conjugate) together with a 
tower of higher spin fields. The scalar is quantised in the standard way. 
At finite $N$ the tower of higher spins is truncated nonperturbatively to a 
maximal spin $N$. This truncation is a consequence  of the equivalence of the 
${\cal W}^{\rm qu}_{\infty}[\mu]$ algebra  to $\W_N$ when $\mu=\frac{N}{N+k}$ and 
$c=c_{N,k}$. This is a bit like a stringy exclusion principle (for a similar truncation in a 
higher spin context see \cite{Ng:2012xp}). This sector is closed under the OPE with itself 
and is a consistent subsector of the theory at large $N$. It is however, not modular invariant 
by itself. For this to be restored we need the nonpertubative states which correspond to the 
scalar dual to $(0;\f)$ and the light states $(\f; \f)$ (or more generally $(\Lambda; \Lambda)$) 
which are nontrivial classical configurations in the bulk --- the analytic continuations of the 
conical defects in the ${\rm SL}(N)$ theory. While it is unusual to have a large number of 
light nonperturbative states, the selection rules of the CFT seem to ensure a good large $N$ 
behavior of correlators \cite{Papadodimas:2011pf, Chang:2011vk}. The upshot seems to be 
that the perturbative Vasiliev theory is highly incomplete at the quantum level and requires 
the various nonperturbative excitations to be taken into account for a consistent completion. 
The novel feature is that these nonperturbative states are not decoupled from the perturbative 
states by virtue of being highly energetic, but rather because of the special nature of the 
interactions --- the fusion rules of the CFT.

\section{Conclusions}

In this paper we have shown how to determine the quantum algebra 
${\cal W}^{\rm qu}_{\infty}[\mu]$ 
underlying the $\hs{\mu}$ higher spin theory on AdS$_3$ explicitly.
In particular, we have managed to find the exact form of the finite $c$
corrections to the commutation relations of the classical 
${\cal W}^{\rm cl}_{\infty}[\mu]$ algebra. This quantum deformation is essentially
uniquely determined  by consistency conditions, in particular the Jacobi identity  
\cite{CGKV}.

While we have not managed to give a closed form
expression for the full quantum algebra, our results are for example sufficient to
determine  the structure of the `minimal' representations exactly. They also
give very strong evidence for the claim that the resulting quantum algebra, 
${\cal W}^{\rm qu}_{\infty}[\mu]$, exhibits an exact `triality' symmetry, relating
in particular the algebra with $\mu=\lambda\equiv \tfrac{N}{N+k}$ to the 
$\W_{N,k}$ minimal model algebra (at finite $N$ and $k$). Given that 
${\cal W}^{\rm qu}_{\infty}[\mu]$ is the only consistent quantum deformation of 
${\cal W}^{\rm cl}_{\infty}[\mu]$, this shows that the symmetries of the
$\W_{N,k}$ minimal model agree with those of the quantum higher spin theory based on
$\hs{\lambda}$. Since the relevant symmetry algebras constrain the theories very
significantly, this goes a long way towards proving the duality at finite $N$ and $k$.

We should stress that the quantum algebra ${\cal W}^{\rm qu}_{\infty}[\mu]$ generically does not
contain $\hs{\mu}$ as a subalgebra, and as a consequence, the representations
of ${\cal W}^{\rm qu}_{\infty}[\mu]$ cannot necessarily be described 
in terms of representations of $\hs{\mu}$. (For example, this is the case for the third
`minimal' representation, see the comments at the end of section~3.2.) The higher spin 
algebra $\hs{\mu}$ only  emerges as a subalgebra for $c\rightarrow \infty$. This is
somewhat reminiscent of the result of Maldacena and Zhiboedov 
\cite{Maldacena:2011jn,Maldacena:2012sf} who showed that, in higher dimensions, 
the higher spin symmetry is necessarily broken by $1/N$ corrections (unless the theory is
free).\footnote{Note that for $\mu=1$, which corresponds to a free theory, 
$\hs{\mu=1} \subset {\cal W}^{\rm qu}_{\infty}[\mu=1]$ is an actual subalgebra, and hence
$\hs{\mu=1}$ remains a genuine symmetry at finite $c$.}
In our case, however, while $\hs{\mu}$ generically is no longer a symmetry of the quantum 
theory, it is replaced by the even larger ${\cal W}^{\rm qu}_{\infty}[\mu]$ algebra
that remains a true symmetry in the quantum theory.

In this paper we have determined the quantum corrections to the classical 
${\cal W}^{\rm cl}_{\infty}[\mu]$ algebra using indirect methods, such as the 
Jacobi identity as well as the representation theory of the ${\cal W}_N$ minimal models.
It would be very interesting to calculate these corrections directly
in the higher spin gravity theory. For example, our analysis makes a specific prediction
for the $1/c$ corrections to the conformal dimension of the perturbative scalar,
see eq.\ (\ref{hlin}), and it would be very interesting to rederive this using perturbation 
theory of the higher spin theory. 

It would also be interesting to study the quantum $\W_\infty$-algebra in 
the supersymmetric case, following 
\cite{Creutzig:2011fe,Candu:2012jq,Henneaux:2012ny,Hanaki:2012yf}; this is currently
under investigation \cite{CGKV}.

\section*{Acknowledgements}

We thank Constantin Candu, Tom Hartman, Maximilian Kelm, Joris Raeymaekers, 
Ashoke Sen and Carl Vollenweider
for useful conversations. The work of MRG is supported in parts by the Swiss National
Science Foundation. RG's research is partially supported by a 
Swarnajayanthi fellowship of the DST, Govt.\ of India, and more broadly by the 
generosity of the Indian people towards basic sciences. MRG thanks 
HRI Allahabad for hospitality during the initial stages
of this work. RG thanks the Isaac Newton Institute, Cambridge for hospitality 
in the final stages. Finally, we both thank the ESI in Vienna for hospitality during the 
penultimate stage.

\appendix

\section{Explicit Commutation Relations}\label{app:comm}

Let us denote the modes of the stress tensor, as usual, by $L_n$, while 
the modes of the spin  $3$, $4$ and $5$ fields are called 
$W_n$, $U_n$ and $X_n$, respectively. Using the ansatz of 
\cite{Gaberdiel:2011wb,GHJ}  and requiring the Jacobi identities 
\be\label{jacobi}
{} [L_m,[L_n,W_l]] + \hbox{cycl.} 
= [L_m,[W_n,W_l]] + \hbox{cycl.} = [U_m,[W_n,L_l]] + \hbox{cycl.} =0 \ ,
\ee
we can determine
the finite $c$ corrections of the commutation relations. The resulting structure
takes then the form
\begin{align}\label{wcomm}
{}[L_m, L_n] &= (m-n)L_{m+n} + \frac{c}{12}m(m^2-1)\delta_{m,-n}\\
{}[L_m, W_n] &= (2m-n)W_{m+n}\\
{}[L_m, U_n] &= (3m-n)U_{m+n}\\
{}[L_m, X_n] &= (4m-n)X_{m+n} \\
{}[W_m, W_n] &= 2(m-n)U_{m+n}+  \frac{N_3}{12}(m-n)(2m^2+2n^2-mn-8)L_{m+n} \\ & \ \ \ \ 
+\frac{8 N_3}{(c+\frac{22}{5})}(m-n)\Lambda^{(4)}_{m+n} 
+ \frac{N_3 c}{144}m(m^2-1)(m^2-4)\delta_{m,-n}\notag\\
{}[W_m, U_n] &= (3m-2n)X_{m+n} - \frac{N_4}{15 N_3}(n^3 - 5m^3 - 3 m n^2 + 5 m^2 n - 9n + 17 m)W_{m+n} \notag \\
& \ \ \ \ 
+ \frac{208 N_4}{25 N_3 (c+\frac{114}{7})}(3m-2n)\, \Lambda^{(5)}_{m+n}
+ \frac{84 N_4}{25 N_3 (c+2)}\, \Theta^{(6)}_{m+n} \ .
\end{align}
The modes of the  composite fields are defined by
\begin{eqnarray}
\Lambda^{(4)}_n & = & \sum_{p} : L_{n-p} L_p :  + \tfrac{1}{5} x_n L_n 
 \\
\Lambda^{(5)}_n & = & \sum_{p} : L_{n-p} W_p : + \tfrac{1}{14} \, y_n W_n \label{L5} \\
\Theta^{(6)}_n & = & \sum_{p} (\tfrac{5}{3} p - n) : L_{n-p} W_p :  
+ \tfrac{1}{6}\, z_n W_n \ , \label{L6}
\end{eqnarray}
where
\begin{eqnarray}
&x_{2l} = (l+1)(1-l) \ , \qquad  & x_{2l-1} = (l+1) (2-l) \ , \\
& y_{2l} = (l+2)(3-5l) \ , \qquad & y_{2l-1} = 5 (l+1)(2-l) \  , \label{ydef} \\
& z_{2l} = l (l+2)  \ , \qquad & z_{2l-1} = 0\ . 
\end{eqnarray}
With this definition we then have the commutation relations
\begin{eqnarray}
{}[L_m, \Lambda^{(4)}_n] & = & (3m-n) \Lambda^{(4)}_{m+n} + 
\bigl( \tfrac{c}{6} + \tfrac{11}{15} \bigr) m (m^2-1) L_{m+n}  \\
{}[L_m, \Lambda^{(5)}_n] & = & (4m-n) \Lambda^{(5)}_{m+n} 
+ \bigl( \tfrac{c}{12} + \tfrac{19}{14} \bigr) m (m^2-1) W_{m+n}  \\
{}[L_m,\Theta^{(6)}_n] & = & (5m-n) \Theta^{(6)}_{m+n} 
+ \bigl( \tfrac{c}{36} + \tfrac{1}{18} \bigr) m (m^2-1) (5m+2n)\,  W_{m+n} \ . 
\end{eqnarray}
The corresponding states are all quasiprimary, and are explicitly given as 
\begin{eqnarray}
\Lambda^{(4)} & = & \bigl( L_{-2} L_{-2}  -\tfrac{3}{5} L_{-4} \bigr) \Omega   \\
\Lambda^{(5)} & = & \bigl( L_{-2} W_{-3}  - \tfrac{3}{7} W_{-5} \bigr) \Omega \\
\Theta^{(6)} & = & \bigl( L_{-3} W_{-3} - \tfrac{2}{3} L_{-2} W_{-4} 
+ \tfrac{1}{2} W_{-6} \bigr) \Omega \ .
\end{eqnarray}
The above commutation relations then satisfy the Jacobi identities (\ref{jacobi}); 
this is true for any value of $N_3$ and $N_4$.
It follows from the classical analysis \cite{Gaberdiel:2011wb} that, to leading order in $1/c$, 
the structure constants take the form
\begin{eqnarray}
N_3 & = & \frac{16}{5}\, q^2 \, (\mu^2-4)  \label{N3c} \\
N_4 & = & \frac{384}{35}\, q^4 \, (\mu^2-4)\,  (\mu^2-9) 
  \ . \label{N4c}
\end{eqnarray} 
Here $q$ is an arbitrary normalisation constant; we can choose it so that $N_3=\tfrac{2}{5}$, 
i.e.\ $q^2 =  \tfrac{1}{8 (\mu^2-4)}$, in which case the OPE of the $W$-field takes the form
\be\label{WWOPE}
W \cdot W \sim \frac{c}{3} \cdot {\bf 1} \  + \ 2 \cdot L \ + \ 
\frac{32}{(5c+22)} \cdot \Lambda^{(4)} \ + \ 4 \cdot U \ .
\ee
Then the leading term in the OPE of $W$ with $U$ equals
\be
W \cdot U \sim \frac{56}{25} \, \frac{N_4}{N_3^2} \, \, W + \cdots = 
\frac{12}{5}\,  \frac{\mu^2-9}{\mu^2-4} \,\, W + \cdots \ .  
\ee
In order to compare this to 
\cite{Hornfeck:1992dp,Hornfeck:1993kp,Blumenhagen:1994wg,Hornfeck:1994is}, 
let us define  
\be\label{betadef}
\hat{U} = \beta^{-1} U \qquad \hbox{with} \qquad 
\beta^2 = \frac{56}{75} \, \frac{N_4}{N_3^2} = \frac{4}{5} \, \frac{\mu^2-9}{\mu^2-4} \ ;
\ee
then the OPEs are of the form
\begin{eqnarray}\label{ww}
W\cdot W & \sim & 
\tfrac{c}{3} \cdot {\bf 1} \ + \ 2 \cdot L \ + \ 8\, \sqrt{\tfrac{1}{5}\, \tfrac{\mu^2-9}{\mu^2-4}} \cdot 
\hat{U} + \cdots \\
W \cdot \hat{U} & \sim & + \ 6\, \sqrt{\tfrac{1}{5}\, \tfrac{\mu^2-9}{\mu^2-4}} \cdot W 
+ \cdots  \ ,
\end{eqnarray}
i.e.\ in the conventions of \cite{Hornfeck:1992tm}, the structure constant $C^4_{33}$ 
equals (\ref{c334c}).

\section{Representation Theory of $\w{\mu}$}\label{app:rep}

In this appendix we want to study the minimal representation of $\w{\mu}$,
whose character is given in eq.~(\ref{minchar}). 

\subsection{Relations at Level One} 

The minimal representation has only a single state at level one, which we may take to be
the $L_{-1}$ descendant of the ground state (which we shall denote by $\phi$). Thus 
we must have the null relations
\begin{eqnarray}
{\cal N}_{1W} & = & \Bigl( W_{-1} - \frac{3w}{2h}L_{-1} \Bigr) \, \phi  \label{13} \\
{\cal N}_{1U} & = & \Bigl( U_{-1} -  \frac{2u}{h} L_{-1} \Bigr) \, \phi  \label{14} \\
{\cal N}_{1X} & = & \Bigl( X_{-1} - \frac{5x}{2h} L_{-1} \Bigr) \, \phi  \label{15} \ .
\end{eqnarray}
Here $w$, $u$ and $x$ are the eigenvalues of the zero modes on $\phi$, i.e.
\be
W_0 \phi = w \phi \ , \qquad
U_0 \phi = u \phi \ , \qquad
X_0 \phi = x \phi \ , 
\ee
and the relative normalisations in ${\cal N}_{1W}$,  ${\cal N}_{1U}$ and  ${\cal N}_{1X}$ are
determined from the condition that $L_1$ annihilates these states.  Actually, if we denote
by $V^{(s)}_n$ the modes of the primary spin $s$ field, then the commutation relations with
the Virasoro algebra take the form
\be
{}[L_m,V^{(s)}_n] = \bigl( (s-1)m - n \bigr) V^{(s)}_{m+n} \ ,
\ee
and hence the corresponding null-vector must be 
\be\label{1gen}
{\cal N}_{1s} = \Bigl( V^{(s)}_{-1} - \frac{ s v(s)}{2h} L_{-1} \Bigr) \phi \ , \qquad \hbox{where} 
\qquad  V^{(s)}_0 \phi = v(s) \phi \ . 
\ee
Again this guarantees that ${\cal N}_{1s}$ is annihilated by $L_1$. Note that (\ref{1gen}) 
generalises the form of the null-vectors (\ref{13}) -- (\ref{15}) to arbitrary spin $s$.

These null-vectors must obviously not just be annihilated by $L_1$, but also by the other
positive modes, i.e.\ by $W_1$, $U_1$, etc., and this will give rise to relations between
the eigenvalues $v(s)$ of the zero modes. For example, from $W_1 {\cal N}_{1W}=0$ we deduce that
\be\label{eq1}
4 u - \frac{N_3}{2} h + \frac{16 N_3}{(c+\frac{22}{5})} \, (h^2+\tfrac{1}{5} h)
- \frac{9}{2} \frac{w^2}{h} = 0 \ ,
\ee
while $W_1 {\cal N}_{1U}=0$  leads to 
\be\label{gq1}
5 x - \frac{4N_4}{5 N_3}  w + \frac{208 N_4}{5 N_3 (c+\frac{114}{7})}w  (h  + \tfrac{3}{7} ) 
- 6 \frac{wu}{h} = 0 \ . 
\ee
Incidentally, this equation can also be obtained from demanding that 
$U_1 {\cal N}_{1W}=0$. 

\subsection{Relations at Level Two}

At level two we may take the linearly independent states to be $L_{-1}^2 \phi$ and $L_{-2}\phi$. 
In particular, we must therefore be able to express $W_{-2}\phi$ in terms of these two states. From
the requirement that the corresponding null-vector must be annihilated by $L_1^2$ and $L_2$ one
finds that it must take the form
\be
{\cal N}_{2W} = \Bigl( W_{-2} + a L_{-1}^2 + b L_{-2} \Bigr) \, \phi \ , 
\ee
with
\be\label{defab}
a = -\frac{3w(2h+c)}{h (16h^2+2ch+c-10h)} \ , \qquad
b = - \frac{24 w (h-1)}{(16h^2+2ch+c-10h)} \ .
\ee
Then we get relations from the requirement that $W_2 {\cal N}_{2W}=0$, and that
$W_1{\cal N}_{2W}$  must be a linear combination of the null-vectors ${\cal N}_{1*}$. The former condition
leads to 
\be\label{eq2}
8 u + 4 N_3 h + \frac{32  N_3}{(c+\frac{22}{5})} \, (h^2 +\tfrac{1}{5} h)+ 12 a w + 6 b w = 0 \ ,
\ee
while the latter condition turns out to require
\be\label{eq3}
\frac{12u}{h} + N_3 + \frac{24 N_3}{(c+\frac{22}{5})} \, (2 h+\tfrac{2}{5}) 
+ a \, \left( 9 \frac{w}{h} + 6 w \right) + \frac{15}{2} b\, \frac{w}{h} = 0 \ .
\ee
Actually, these relations are linearly dependent, and thus 
we cannot determine all eigenvalues directly, but we can express both 
$w$ and $u$ as functions of $h$; indeed we can eliminate $u$ by combining 
(\ref{eq1}) and (\ref{eq2}), and then  obtain 
\be\label{wh}
w(h) = \pm \frac{h}{3} \,  \sqrt{ \frac{- 5 N_3\,  (16 h^2+2ch+c-10h)} {(2ch-3c-2h)}} \ .
\ee
Similarly, we can determine $u\equiv u(h)$ as 
\be\label{uh}
u = - h\, N_3  \; \frac{c^2 - 12 c h  + c^2 h + 18 h^2 c + 2c + 36 h^2-28 h}{(c+\frac{22}{5}) 
\, (2 c h - 3 c - 2 h) } \ .
\ee
Once we have these relations, we can then determine all eigenvalues $v(s)$ recursively as a function of $h$. 
To this end we consider the relations coming from $W_1 {\cal N}_{1s}=0$ for $s=4,5,6,\ldots$. For example,
for $s=4$, this is just (\ref{gq1}), which we can solve for $x$ as 
\be\label{reln1}
x = w \, \Bigl(   \frac{4N_4}{25 N_3}  - \frac{208 N_4\, }{25 N_3 (c+\frac{114}{7})}  (h+\tfrac{3}{7})
+ \frac{6 u}{5 h} \Bigr) \ .
\ee
Since we already know $w$ and $u$ as a function of $h$, this therefore leads to an expression for
$x\equiv x(h)$. For the general case, we note that the OPE of the spin $3$ field $W$ with $V^{(s)}$ 
will only involve simple fields of spin at most $s+1$. Thus $W_1 {\cal N}_{1s}=0$ will lead to a 
relation between $v(s+1)$ and $v(t)$ with $t\leq s$. Recursively, we can therefore determine all 
$v(s)$ in terms of $h$. 

\subsection{The Final Equation}

Thus it only remains to find one final equation which will allow us to also determine $h$. To find this
equation we now evaluate the condition $U_2 {\cal N}_{2W} = 0$, which leads to 
\be\label{reln2}
10 x + 4 w \frac{N_4}{N_3} + \frac{416 N_4}{5 N_3 (c+\frac{114}{7})} 
w\, (h +\tfrac{3}{7}) + 20 a u + 8 b u = 0 \ .
\ee
Instead of demanding that $U_2 {\cal N}_{2W}=0$ we may also study the condition 
that $U_1{\cal N}_{2W}$ is a linear combination of the null-vectors at level one, i.e.\ 
the vectors ${\cal N}_{1*}$ given in (\ref{13})-(\ref{15}). This leads to 
\be\label{reln3}
10x + w  \frac{6N_4}{5 N_3}  +  \frac{416 N_4\, (h+\tfrac{3}{7})}{5 N_3 (c+\frac{114}{7}) }  w 
+ 12a u + 4a u h +7 b u  = 0 \ .
\ee
Together with (\ref{gq1}) these three equations are indeed linearly dependent.
In order to solve for $h$, we now equate (\ref{reln1}) with (\ref{reln3}) to obtain
\be\label{reln5}
4 a \frac{u}{w} (3 + h) + 7 b \frac{u}{w}   +\frac{12 u }{h} + \frac{14N_4}{5 N_3} =0 \ ,
\ee
where $a$ and $b$ are defined as in (\ref{defab}). 
Using the expressions for  $u$ from (\ref{uh}) we then 
get the cubic equation for $h$ given in eq.~(\ref{hcube}).

\section{The Fusion of $(\f;\f)$ at Finite $N$ and $c$}\label{app:fus}

The fusion analysis of $(\f;0)\otimes (0;\f)$ can essentially be done
using the steps described in \cite{Gaberdiel:2011zw}, so we shall be somewhat brief and
only stress the main differences. Let us denote the highest weight
states of the relevant representations by $\phi_1=(\f;0)$ and
$\phi_2=(0;\f)$. Following the discussion of appendix~\ref{app:rep}, 
we then have the null-vectors 
\be
\begin{array}{rclrcl}
\bigl( W_{-1} - \frac{3w_1}{2h_1} L_{-1} \bigr) \phi_1 & = & 0 \qquad
&\bigl( W_{-1} - \frac{3w_2}{2h_2} L_{-1} \bigr) \phi_2 & = & 0  \\[4pt]
\bigl( W_{-2} +a_1 L_{-1}^2 + b_1 L_{-2} \bigr) \phi_1 & = & 0 \qquad
& \bigl( W_{-2} +a_2 L_{-1}^2 + b_2 L_{-2} \bigr) \phi_2 & = & 0 \ ,
\end{array}
\ee
where $a_j=a(h_j,w_j)$ and $b_j=b(h_j,w_j)$, $j=1,2$, are defined by (\ref{defab}), and
in the parametrisation of section~\ref{finiteNsub}, the eigenvalues equal (for $N_3=\tfrac{2}{5}$) 
\be
\begin{array}{rclrcl}
h_1 & = & \tfrac{(N-1)(2N+k+1)}{2N (N+k)} \qquad
& w_1 & = & - \, \tfrac{\sqrt{2} (N-1) (2N+k+1) }{6 N (N+k)}\,
 \sqrt{\tfrac{(N-2)(3N+2k+2)}{N (N+2k)}} \\[8pt]
h_{2} & = & \tfrac{(N-1) k}{2 N (N+k+1)} \qquad
& w_2 & = & \tfrac{\sqrt{2} (N-1) k }{6N  (N+k+1)} \, 
\sqrt{\tfrac{(N-2) (N+2k)}{N(3N+2k+2)}} \ .
\end{array}
\ee
Note that in the 't~Hooft limit, $N,k\rightarrow \infty$ with $\lambda=\tfrac{N}{N+k}$, 
we have the familiar relations
\be
\begin{array}{lcrlcr}
h_1 & \simeq &  \tfrac{1}{2} (1 + \lambda) \qquad
w_1 & \simeq & - \tfrac{\sqrt{2}}{6} \, (1+\lambda) \, \sqrt{\frac{(2+\lambda)}{(2-\lambda)}}  \\[8pt]
h_2 & \simeq &  \tfrac{1}{2} (1 - \lambda) \qquad
w_2 & \simeq &  \tfrac{\sqrt{2}}{6}\,  (1-\lambda) \, \sqrt{\frac{(2-\lambda)}{(2+\lambda)}}   \ . 
\end{array}
\ee
Furthermore, the parameters $a_j$ and $b_j$ simplify in that limit to 
\be\label{ablim}
a_j \simeq \tfrac{3 w_j}{h_j(2 h_j+1)} = (-1)^j\, \sqrt{\tfrac{2}{(4-\lambda^2)}} \ , \qquad
b_j \simeq 0 \ ,
\ee
since $c\rightarrow \infty$ in that limit.

Let us study the highest weight space of the fusion product. Then we have the relations
\begin{eqnarray}
0 \cong   \Delta(W_{-1}) & = & (W_{-2}\otimes {\bf 1}) + (W_{-1}\otimes {\bf 1}) 
+ ({\bf 1}\otimes W_{-1}) \\
& = & - a_1 (L_{-1}^2 \otimes {\bf 1}) - b_1 (L_{-2} \otimes {\bf 1}) 
+ \tfrac{3w_1}{2h_1} (L_{-1} \otimes {\bf 1})  +  \tfrac{3w_2}{2h_{2}} ({\bf 1}\otimes L_{-1} ) \nonumber  \\
& = & - a_1 (L_{-1}^2 \otimes {\bf 1}) 
+ \bigl( \tfrac{3w_1}{2h_1} - \tfrac{3w_2}{2h_2} + b_1 \bigr) (L_{-1} \otimes {\bf 1}) 
- b_1 h_2  \, ({\bf 1} \otimes {\bf 1})\ , \nonumber
\end{eqnarray}
where we have used that on the highest weight space
\begin{eqnarray}
({\bf 1} \otimes L_{-1}) & \cong & - (L_{-1} \otimes {\bf 1}) \label{eeq1} \\
(L_{-2}\otimes {\bf 1}) & \cong & ({\bf 1}\otimes L_{-1}) + ({\bf 1}\otimes L_{0}) \ . \label{eeq2} 
\end{eqnarray}
Thus we obtain the identity
\be\label{id1}
a_1 (L_{-1}^2 \otimes {\bf 1})  = 
\bigl( \tfrac{3w_1}{2h_1} - \tfrac{3w_2}{2h_2} + b_1 \bigr) (L_{-1} \otimes {\bf 1}) 
- b_1 h_2 \, ({\bf 1} \otimes {\bf 1}) \ .
\ee
Incidentally, in the 't~Hooft limit, this reduces to the identity
$(L_{-1}^2\otimes {\bf 1}) \cong - 2 (L_{-1}\otimes {\bf 1})$, 
see eq.~(5.17) of \cite{Gaberdiel:2011zw}, since $b_j\rightarrow 0$ in that limit, see
(\ref{ablim}).

\noindent A second identity can be obtained from 
\begin{eqnarray}
0 \cong  - \Delta(W_{-2}) & = & - (W_{-2}\otimes {\bf 1}) - ({\bf 1}\otimes W_{-2}) \\
& = &  a_1 (L_{-1}^2 \otimes {\bf 1}) + b_1 (L_{-2} \otimes {\bf 1}) + 
a_2 ({\bf 1}\otimes L_{-1}^2 ) + b_2 ({\bf 1} \otimes L_{-2})  \nonumber \\
& = & (a_1+a_2) (L_{-1}^2 \otimes {\bf 1}) - (b_1+b_2) (L_{-1} \otimes {\bf 1}) 
+(b_1h_2 + b_2 h_1) \, ({\bf 1} \otimes {\bf 1}) \ , \nonumber
\end{eqnarray}
where we have used (\ref{eeq1}), (\ref{eeq2}) as well as 
\be
({\bf 1} \otimes L_{-2}) \cong - (L_{-1} \otimes {\bf 1}) + (L_0 \otimes {\bf 1}) \ .
\ee
Thus we obtain a second identity, namely
\be\label{id2}
(a_1+a_2) (L_{-1}^2 \otimes {\bf 1}) =  (b_1+b_2) (L_{-1} \otimes {\bf 1}) 
- (b_1h_2 + b_2 h_1) \, ({\bf 1} \otimes {\bf 1}) \ .
\ee
Note that this identity becomes trivial in the 't~Hooft limit 
(since $b_j\rightarrow 0$ and $a_1\rightarrow - a_2$, see (\ref{ablim})), but for
finite $c$, we can deduce from  (\ref{id1}) and (\ref{id2}) the relation 
\be
 \Bigl[ \bigl( \tfrac{3w_1}{2h_1} - \tfrac{3w_2}{2h_2} \bigr) (a_1+a_2) 
+ \bigl(b_1 a_2 - b_2 a_1\bigr) \Bigr] (L_{-1} \otimes {\bf 1})  \cong
\bigl(  a_2 b_1 h_2 - a_1 b_2 h_1 \bigr) ({\bf 1} \otimes {\bf 1}) \ ,
\ee
which simplifies to 
\be
(L_{-1}\otimes {\bf 1}) \cong - \tfrac{N-1}{N} \, ({\bf 1} \otimes {\bf 1}) \ . 
\ee
Thus for finite $c$, the highest weight space is one-dimensional, and the $L_0$
eigenvalue of $({\rm f},{\rm f})$ becomes indeed just (\ref{fusion}).
\smallskip

One can also determine the $W_0$ eigenvalue of the ground state of the fusion
product; using 
\be
\Delta(W_0) = (W_{-2} \otimes {\bf 1}) + 2 (W_{-1} \otimes {\bf 1}) + (W_0 \otimes {\bf 1})
+ ({\bf 1}\otimes W_0)
\ee
one finds that 
\be\label{wff}
\begin{array}{rcl}
w(\f;\f) & = & w_1 + w_2 - \frac{N-1}{N} \Bigl( \frac{3w_1}{2h_1} + \frac{3w_2}{2h_2} \Bigr) \\
& = & - \sqrt{\frac{2(N-2)}{N}}  \, 
\frac{(N^2-1)(N+2)}{6N} \, \frac{1}{(N+k)(N+k+1)}\, \sqrt{\frac{1}{(N+2k)(3N+2k+2)}} \ .
\end{array}
\ee
In order to bring this into the form of (\ref{eigenvals}), we have to rescale $w$, 
i.e.\ we have to work with
\be
\hat{N}_3=\frac{2}{5}\, ( (N+2) c + (3N+2)(N-1)) = \frac{2}{5}\, 
\frac{(N^2-1)\, (N+2k)(3N+2k+2)}{(N+k) (N+k+1)} 
\ee
instead. Then the corresponding $\hat{w}$ eigenvalue equals
\be\label{wfffin}
\hat{w}(\f;\f) =  - \sqrt{\frac{2(N-2) (N^2-1)}{N}}  \, 
\frac{(N^2-1)(N+2)}{6N} \, \Bigl(\frac{1}{(N+k)(N+k+1)}\Bigr)^{\frac{3}{2}} \ ,
\ee
and is hence of the form  (\ref{eigenvals}).

\bibliographystyle{JHEP}

\end{document}